\documentclass[twocolumn,aps,prl,showpacs,superscriptaddress]{revtex4-1}
\usepackage[latin9]{inputenc}
\usepackage{amsmath}
\usepackage{amssymb}
\usepackage{graphicx}

\usepackage{graphics}
\usepackage{epsfig}
\usepackage{epstopdf}
\usepackage{slashed}

\usepackage[usenames,dvipsnames]{xcolor}
\definecolor{linkcolor}{RGB}{55,57,154}
\usepackage[pdfstartview={XYZ},colorlinks=true,allcolors=linkcolor,pdffitwindow=false,bookmarksopen=true]{hyperref} 

\IfFileExists{srcltx.sty}{\usepackage[active]{srcltx}}

\newcommand{\be}{\begin{equation}}
\newcommand{\ee}{\end{equation}}
\newcommand{\bea}{\begin{eqnarray}}
\newcommand{\eea}{\end{eqnarray}}

\makeatother

\begin{document}

\title{Postinflationary Higgs Relaxation and the Origin of Matter-Antimatter Asymmetry}

\author{Alexander Kusenko}
\affiliation{Department of Physics and Astronomy, University of California, Los Angeles, California 90095-1547, USA}
\affiliation{Kavli Institute for the Physics and Mathematics of the Universe (WPI), University of Tokyo, Kashiwa, Chiba 277-8568, Japan}

\author{Lauren Pearce}
\affiliation{William I.\ Fine Theoretical Physics Institute, School of Physics and Astronomy, University of Minnesota, Minneapolis, Minnesota 55455, USA}

\author{Louis Yang}
\affiliation{Department of Physics and Astronomy, University of California, Los Angeles, California 90095-1547, USA}

\preprint{FTPI-MINN-14/32}


\begin{abstract}
The recent measurement of the Higgs boson mass implies a relatively slow rise of the standard model Higgs potential at large scales, and a possible second minimum at even larger scales.  Consequently, the Higgs field may develop a large vacuum expectation value during inflation.  The relaxation of the Higgs field from its large postinflationary value to the minimum of the effective potential represents an important stage in the evolution of the Universe. During this epoch, the time-dependent Higgs condensate can create an effective chemical potential for the lepton number, leading to a generation of the lepton asymmetry in the presence of some large right-handed Majorana neutrino masses.  The electroweak sphalerons redistribute this asymmetry between leptons and baryons.  This Higgs relaxation leptogenesis can explain the observed matter-antimatter asymmetry of the Universe even if the standard model is valid up to the scale of inflation, and any new physics is suppressed by that high scale.

\end{abstract}
\maketitle

The recent discovery of a Higgs boson with mass 125~GeV~\cite{Chatrchyan:2012ufa,Aad:2012tfa} implies that the Higgs potential is very shallow and may even develop a second minimum, assuming that the standard model is valid at high energy scales~\cite{Degrassi:2012ry}. During cosmological inflation, the Higgs field may be trapped in a quasistable second minimum or, alternatively, may develop a stochastic distribution of vacuum expectation values due to the flatness of the potential~\cite{Bunch:1978yq,Linde:1982uu,Starobinsky:1994bd}.  In both scenarios, the Higgs field relaxes to its vacuum state after inflation via a coherent motion.  In this Letter we explore this epoch of Higgs relaxation. 

We show that the observed matter-antimatter asymmetry could arise during this epoch.  The Sakharov conditions~\cite{Sakharov:1967dj}, necessary for baryogenesis, 
are generically satisfied in the presence of the out-of-equilibrium Higgs condensate evolving with time~\cite{Cohen:1987vi,Dine:1990fj} 
and the neutrino Majorana masses that violate the lepton number.  

The standard model Higgs boson has a tree-level potential $V(\Phi)= m^{2}\Phi^{\dag}\Phi+\lambda(\Phi^{\dag}\Phi)^{2}$, where $\Phi$ is an SU(2) doublet. Using a gauge transformation, one can write the classical field as $\Phi=\left(1\slash\sqrt{2})\{e^{i\theta}\phi,0\right\} $, where $\phi(x)$ is real.  Loop corrections substantially modify this potential at large values.  We will include one-loop and finite temperature corrections to the Higgs potential, although two-loop effects may also be important near the metastability boundary~\cite{Degrassi:2012ry}. For the experimentally preferred top and Higgs mass values, the $\sqrt{\langle\phi^2\rangle}=v_{\rm EW}=246$~GeV minimum appears to be metastable~\cite{Degrassi:2012ry}, which entails a number of important ramifications~\cite{unstable}.  However, a stable vacuum is still possible within the experimental uncertainties~\cite{Degrassi:2012ry}. Furthermore, higher-dimensional operators can modify the potential at large vacuum expectation value (VEV)~\footnote{Here and below, we identify the VEV with the expectation value of $\sqrt{\protect\langle \phi^2\protect\rangle}$, while $\protect\langle \phi\protect\rangle=0$.}  and make the vacuum stable. During inflation, a scalar field may develop a nonzero VEV $\left<\phi^2 \right>$ for more than one reason.  We will consider two cosmological scenarios that lead to two types of initial conditions.

{\em Initial condition 1 (IC-1)}.---IC-1 occurs for the central values of measured Higgs and top quark masses.  A false vacuum can appear at large $\phi$ due to the negative effective coupling  $\lambda_{\rm eff}(\phi)$ in the potential, stabilized by some higher-dimensional operator(s). If the VEV prior to inflation is large (similar to the initial VEV of the inflaton in chaotic inflation), then the field can evolve toward the false vacuum from above and may become trapped in the false vacuum.   When reheating begins, finite-temperature effects eliminate this minimum, and the field rolls down to the global minimum at $\left< \phi^2 \right> = 0$.  

{\em Initial condition 2 (IC-2).}---During inflation, scalar fields with slowly rising potentials develop large VEVs~\cite{Bunch:1978yq,Linde:1982uu,Starobinsky:1994bd,Lee:1985uv,Affleck:1984fy}.   The qualitative reason is that, in a de~Sitter space, a scalar field can perform a quantum jump via a Hawking-Moss instanton~\cite{Bunch:1978yq,Hawking:1981fz}.  The subsequent relaxation by means of a classical motion requires time of the order $\tau_{\phi}\sim\sqrt{d^{2}V/d\phi^{2}}$. If the Hubble parameter during inflation, $H_{I}=\sqrt{8\pi/3}\Lambda_{I}^{2}/M_{P}$, is much greater than $\tau_\phi$, then relaxation is too slow and quantum jumps occur frequently enough to maintain a large VEV. Averaged over superhorizon scales, the mean Higgs VEV is such that $V(\phi_I)\sim H_I^{4}$~\cite{Bunch:1978yq,Hawking:1981fz,Enqvist:2013kaa}.  This scenario occurs if the standard Higgs vacuum is stable (which is consistent with the Higgs and top mass measurements, although not with the central values), or if the $\left< \phi^2 \right> = 0$ minimum is quasistable and the scale of inflation is not sufficiently high to probe the false vacuum.

We will see that the Higgs relaxation in both cases (IC-1 and IC-2) can explain the baryon asymmetry of the Universe, and the asymmetry depends on the initial value of the VEV, denoted by $\sqrt{\left< \phi^2 \right>} \equiv \phi_0$.  As quantum fluctuations of the Higgs field were ongoing during inflation, different patches of the Universe began with slightly different $\phi_0$ values and consequently developed different baryon asymmetries.  This could result in unacceptably large baryonic isocurvature perturbations~\cite{isocurvature}, which are constrained by cosmic microwave background observations~\cite{Ade:2013uln}.  These constraints can be satisfied as follows.

In the case of IC-1, we make use of the additional term that stabilizes the second minimum; we choose a term that also ensures $m_\mathrm{eff} > H_I$ in the false vacuum. As a concrete example, one such term for the experimentally preferred values $m_h = 126$~GeV and $m_t = 173.07$~GeV is
\begin{equation}
\mathcal{L}_\mathrm{lift} =  \dfrac{(\phi^\dag \phi)^{5}}{(6.52 \times 10^{15} \, \mathrm{GeV})^{6}}.
\label{eq:lift}
\end{equation}
In this scenario, the parameters must be chosen such that the barrier is large enough that the Hawking-Moss instanton does not destabilize the false vacuum during inflation, but the reheat temperature does destabilize the vacuum.

In the IC-2 case, we consider a coupling between the Higgs field and the inflaton via one or several operators of the form 
\begin{equation}
\mathcal{L}_\mathrm{\phi I} = c \dfrac{(\phi^\dagger \phi)^{m/2} (I^\dagger I)^{n \slash 2}}{M_{P}^{m+n-2}},
\label{eq:inflaton_coupling}
\end{equation}
which increases the effective mass of the Higgs field during inflation. While $\langle I \rangle $ is large (super-Planckian, in the case of chaotic inflation), 
this term limits the Higgs VEV and can be chosen to give  $m_\mathrm{eff}(\phi_I) \gg H_I$ during most of the inflationary epoch.  During the last $N_\mathrm{last}$ $e$-folds of inflation, the inflaton VEV is small enough for the Higgs VEV to grow to some value $\phi_0=\min [\phi_I, \sqrt{N_\mathrm{last}} H_I \slash 2 \pi]$. If $N_\mathrm{last}$ is sufficiently small, the baryonic isocurvature perturbations develop only on the smallest angular scales, on which they have not yet been constrained.  Since the change in $\langle I \rangle$ during the slow-roll phase of inflation is model dependent, the choice of parameters $c,m,n$ differs from model to model and may require some fine-tuning.  

We note that both operators \eqref{eq:lift} and \eqref{eq:inflaton_coupling} may be viewed as effective operators arising from loops after integrating out heavy states. 
We also note that, although Higgs relaxation commences at the early stages of reheating, the energy density is never dominated by the Higgs field.  
Inflaton oscillations dominate until the transition to the radiation dominated era, which occurs when the inflaton decay width $\Gamma_{I}\sim H_{\rm RH}$, 
at a much later time~\cite{Abbott:1982hn}. The resulting reheat temperature $T_{\rm RH}\sim\sqrt{\Gamma_{I}M_{P}}\ll H_{I}$ is allowed over a broad range of values~\cite{Dai:2014jja}.

The large VEV of the Higgs field makes its dynamics sensitive to higher-dimensional operators which are normally suppressed by some power of a high scale. A number of different operators can play an important role in Higgs relaxation. We will consider the following operator, familiar from spontaneous baryogenesis models~\cite{Dine:1990fj}:
\begin{equation}
{\cal O}_{6}  =  -\frac{1}{M_{n}^{2}}(\partial_{\mu}|\phi|^{2})\, j^{\mu}.
\end{equation}
$j^{\mu}$ is the fermion current whose zeroth component is the density of $(B+L)$.  This operator is equivalent to
\begin{equation}
\mathcal{O}_6 = -\dfrac{3}{16 \pi^2 M_{n}^{2}} |\phi^2| \left( g^2 W \tilde{W} - g^{\prime 2} \dfrac{1}{2} A \tilde{A} \right),
\end{equation}
[where $W$ and $A$ are the the $\mathrm{SU}_L(2)$ and $\mathrm{U}_Y(1)$ gauge fields, respectively] through the mixed SU(2)$\times$U(1) anomaly~\cite{Dine:1990fj}.  A term of this form may be generated through a loop with fermions which couple to $\mathrm{SU}_L(2)$ vectorially and have soft mass terms with $CP$-violating phases, as Higgsinos and gauginos in supersymmetric models.  We also note that thermal loops can produce a similar term with $T$ in place of $M_n$~\cite{Shaposhnikov:1987tw,Shaposhnikov:1987pf,Smit:2004kh,Brauner:2012gu}.  Provided that the temperature evolution is slow with respect to the time evolution of the Higgs VEV, one may approximate $\partial_t (|\phi(t)|^2\slash T(t)^2) \approx (\partial_t |\phi|^2) \slash T^2$.

While the Higgs VEV $\phi(t)$ is in motion, the Lagrangian contains the term $(-\mu_{\textrm{eff}}\, j_{B+L}^{0})$, where  
\begin{eqnarray}
 \mu_{\textrm{eff}}=
\partial_{t}|\phi|^{2}\slash M_{n}^{2}. 
\label{eq:mu}
\end{eqnarray}
This term spontaneously breaks charge, parity, and time reversal (CPT) symmetry~\cite{Cohen:1987vi}, and acts as a chemical potential, shifting the energy levels of leptons as compared to antileptons. Lepton number violating processes allow the system to lower its free energy at some value of $(B+L)\neq 0$. Since $(\partial_{t}|\phi|^{2})$ changes sign during the oscillations of the Higgs VEV $\phi$, there is partial cancellation during the oscillation of the Higgs VEV, but this cancellation is not exact due to the decrease in the amplitude of the VEV.  The sign of the final asymmetry is determined by the first, large swing of the field.  This sign is the same everywhere in the Universe because $|\phi|^2$ decreases with time and $\partial_t|\phi|^2$ is negative. 

\begin{figure}[ht!]
\includegraphics[width=0.9\columnwidth]{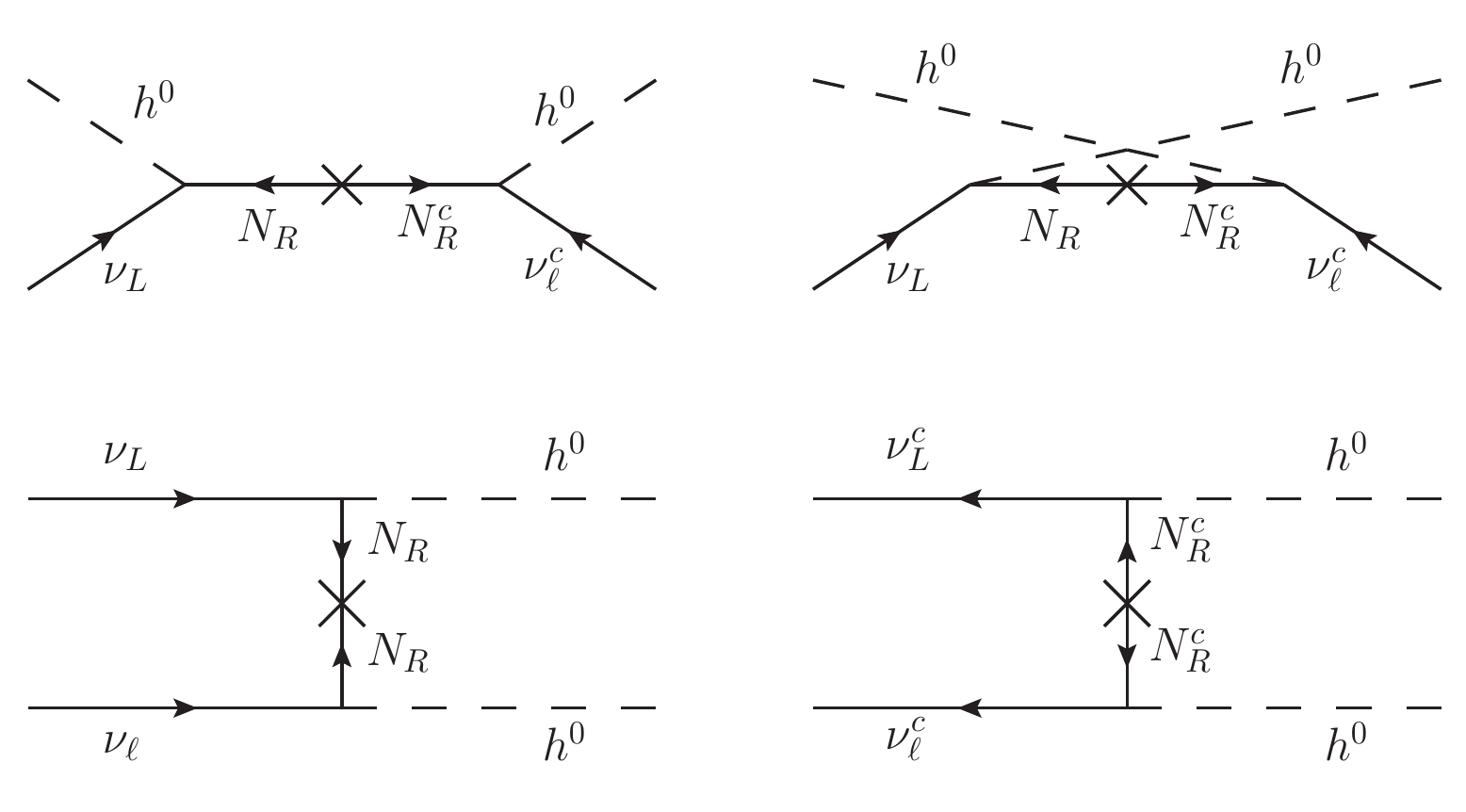}
\protect\caption{\label{fig:diagram} Some diagrams that contribute to lepton number violation via exchange of a heavy Majorana neutrino.}
\end{figure}

We assume the standard seesaw~\cite{seesaw} mass matrix for neutrinos and require that the Majorana mass $M_{R}\gg T, M_R \gg m_\mathrm{eff}(\phi_{0})$ to forbid 
production of right-handed neutrinos in thermal plasma and in the condensate decay.  The Majorana mass allows for processes violating the lepton number $L$ [and, therefore, $(B+L)$]. 
Such processes involve an exchange of virtual heavy Majorana neutrinos; some of them are shown in  Fig.\ \ref{fig:diagram}.
In the presence of Majorana mass terms, the effective lepton number $L$ is the sum of the lepton numbers of the charged leptons and the helicities of the light neutrinos. This is conserved in the limit $M_{R}\rightarrow\infty$, but not conserved for a finite $M_R$.

These lepton number violating processes change the density of the $(B-L)$ charge, which is conserved in all other processes, including the sphaleron transitions. At the same time, the U(1) symmetry corresponding to $(B+L)$ is anomalous, and the sphaleron transitions can change this quantity at a characteristic rate per unit volume $\sim(\alpha_{W}T)^{4}\exp\{-g_{W}\phi(t)/T\}$. During the relaxation of the Higgs VEV, this rate can change dramatically from a slow rate, when $B$ and $L$ are conserved separately (for a large VEV), to a rapid rate, at which $B$ and $L$ densities approach the equilibrium values such that $n_{B}=(28/79)n_{B-L}$.  However, once the asymmetry in $L$ develops, the sphaleron transitions do not change $L$ by more than a factor of order 1. Therefore, it is appropriate to concentrate on either the lepton number or on $(B-L)$ for order-of-magnitude estimates.

We note that, depending on the value of $\phi(t)$, weak interactions, mediated by heavy weak bosons with masses $M_{W}\propto\phi(t)$, may be in or out of equilibrium in the plasma created by inflaton decay.  When $\phi(t)$ is close to zero, weak interactions equilibrate the distributions of charged and neutral leptons.  Neutrinos and Higgs bosons may also be produced by the decay of the inflaton directly.  For definiteness, we assume a thermal number density for each of these species.  Consequently, the rate of lepton number violating [and $(B-L)$ violating] processes per unit volume due to the diagrams shown in Fig.~\ref{fig:diagram} can be estimated as $(\Gamma_{\slashed{L}}/{\rm Vol.})\sim T^{6}\sigma_{R}$, where $\sigma_{R}$ is the thermally averaged cross section of the interactions involving the heavy neutrino exchange.  The exact rate depends on the temperature of the plasma and the flavor structure in the Yukawa matrix $Y_{ij}$ that enters into the neutrino mass matrix. However, one can estimate 
\begin{equation}
\sigma_{R}\simeq\frac{|(Y^{\dag}Y)_{jj}|^2}{4 \times 16\pi M_{R,j}^{2}}\simeq\frac{\sum_{j}m_{\nu,j}^{2}}{16\pi v_{\rm EW}^{4}}\sim10^{-31}\,{\rm GeV}^{-2},
\label{eq:cross_section}
\end{equation}
where $M_{R,j}$ are the right-handed neutrino masses and the reheat temperature is assumed to be too low to produce on-shell right-handed neutrinos.  Since the right-handed neutrinos are not present in plasma and not produced via Higgs condensate decay, the contribution from standard leptogenesis~\cite{Fukugita:1986hr} is strongly suppressed. [We note that Eq.~\eqref{eq:cross_section} neglects several $\mathcal{O}(1)$ factors, most notably from thermal averaging and the resonance in the $s$-channel diagram.]

Based on detailed balance, one can describe the evolution of the particle number densities by a system of Boltzmann equations. One expects the lepton number density $n_{L}=n_{\nu}-n_{\overline{\nu}}$ of each species to relax to its equilibrium value $\sim\mu_{\mathrm{eff}}T^{2}$ linearly, which gives the approximate equation 
\begin{equation}
\frac{d}{dt}n_{L}+3Hn_{L}\cong-\frac{2}{\pi^{2}}T^{3}\sigma_{R}\left(n_{L}-\frac{2}{\pi^{2}}\mu_{\textrm{eff}}\, T^{2}\right).
\label{eq:nL}
\end{equation}

The effective temperature of the plasma during reheating is defined through its radiation density, which comes from decays of the inflaton $I$ and evolves with time as~\cite{Kolb:1990vq}
$\rho_{R}=(g_{*}\pi^{2}/30)T^{4}\simeq(M_{P}^{2}\Gamma_{I}/10\pi)\left[1/(t+t_{i})\right]\{1-\left[t_{i}/(t+t_{i})\right]^{5/3}\}$,
where $t_{i}=(2/3)\sqrt{3\slash8\pi}M_{{\rm P}}\slash\Lambda_{I}^{2}$
and $t=0$ corresponds to the start of coherent oscillations of the
inflaton field.  For $t \gg t_i$, the temperature evolves as
\begin{equation}
T=\left( \frac{3}{g_*\pi^3}\frac{\Gamma_I M_P^2}{t}\right)^{1/4},
\label{eq:time1}
\end{equation}
until it reaches the reheat temperature $T_R\sim \sqrt{\Gamma_I M_P}$, at which point the radiation dominates the energy density.  After this the temperature evolves as 
\begin{equation}
 T=\left( \frac{45}{16\pi^3 g_*}\right)^{1/4} \sqrt{M_P/t} .
 \label{eq:time2}
\end{equation}

The approximate equation \eqref{eq:nL} can be analyzed in two regimes: during the relaxation of the Higgs VEV [$\mu_{\rm eff}(t) \neq 0$] and during the subsequent cooling of the Universe ($\mu_{\rm eff}=0$). During the Higgs relaxation, which occurs on the time scale of the order of $H_{\phi}^{-1}$, 
\begin{eqnarray}
 \mu_{{\rm eff}} = 
\frac{\partial_{t}|\phi^{2}|}{M_{n}^{2}}\sim\frac{H_{\phi}\phi_{0}^{2}}{M_{n}^{2}}\sim\frac{\sqrt{\lambda}\phi_{0}^{3}}{M_{n}^{2}}. 
\end{eqnarray}
As the Higgs VEV oscillates, the equilibrium value is 
\begin{equation}
n_{L,{\rm eq}}\sim\mu_{{\rm eff}}T^{2}\sim\frac{\sqrt{\lambda}\phi_{0}^{3}T_{{\rm max}}^{2}}{M_{n}^{2}}\sim 
\frac{\sqrt{\lambda}\phi_{0}^{3}
T_R \Lambda_I}{M_{n}^{2}}
\end{equation}
However, the relevant reactions may not be fast enough to equilibrate to this value before the Higgs VEV approaches zero at $t_{\textrm{rlx}}$. 
In this case, the maximum asymmetry reached by the end of Higgs relaxation at time $t_{\textrm{rlx}}$ is $n_{L,\textrm{eq}} \sigma_{R} T_{\textrm{rlx}}^{3} t_{\textrm{rlx}}$, where $T_{\textrm{rlx}}$ is the temperature at $t_{\textrm{rlx}}$.  In either case, 
\begin{equation}
n_{\textrm{rlx}}\sim n_{L,\textrm{eq}}\times \min \left \{ 1, (\sigma_{R} T_{\textrm{rlx}}^{3} t_{\textrm{rlx}}) \right \}.
\label{eq:initial}
\end{equation}

After the Higgs relaxation is completed at $t_{{\rm rlx}}$, the generated lepton asymmetry can be partially washed out by heavy neutrino exchanges, until these go out of equilibrium.  During washout, Eq.~\eqref{eq:nL} can be rewritten as 
\begin{equation}
\frac{dN_{L}(t)}{dt}\simeq-\frac{2}{\pi^{2}}T^{3}\sigma_{R}N_{L}(t),
\label{eq:comoving}
\end{equation}
where $N_{L}\equiv n_{L}a^{3}$ is the lepton number per comoving volume.  Using the functions $T(t)$ from Eqs.~\eqref{eq:time1} or \eqref{eq:time2}, one can rewrite 
and solve Eq.~\eqref{eq:comoving} for the evolution of $N_L$ as a function of temperature $T$: 
\begin{equation}
 N_L(T)= 
 \begin{cases}
 N_L(T_{\rm rlx}) \exp\left[-\frac{24(\sigma_R M_P^2) \Gamma_I}{g_* \pi^5}  (T^{-1}-T_{\rm rlx}^{-1}) \right], \\ \hfill T \ge T_R \\
 N_L(T_{R}) \exp\left[-\frac{3\sqrt{5}\sigma_R M_P}{\sqrt{g_* \pi^7}}  (T_R-T) \right], T < T_R \\
  \end{cases},
\end{equation}
provided that $t_\mathrm{rlx} \gg t_i$. At the end of reheating,
\begin{align}
 N_L(T_R)= N_L(T_{\rm rlx}) \exp\left[-\frac{24}{g_* \pi^5} \sqrt{ \dfrac{g_* \pi^3}{3}} \sigma_R M_P T_R\right],
\end{align}
where we set $T=T_R$, assumed that $T_{\rm rlx}\gg T_R$, and used the relation $\Gamma_I/T_R=T_R/M_P \sqrt{ g_*  \pi^3 \slash 3 }$. The asymptotic value at low temperature is 
\begin{align}
 N_L(T \rightarrow 0)=N_L(T_{\rm rlx}) \exp\left[-\left( \frac{24 + 3\sqrt{15}}{\sqrt{ 3 g_* \pi^7}}\right) \sigma_R M_P T_R\right].
\end{align}
To calculate the asymmetry we include the dilution by entropy production from the time $t_{\rm rlx}$ to the time of reheating.  The comoving entropy is conserved for $T < T_R$ until the standard model degrees of freedom go out of equilibrium.  Therefore, taking into account the dilution by factor 
$(a_\mathrm{rlx} \slash a_R)^3 \approx (t_\mathrm{rlx} \slash t_R)^2 = t_\mathrm{rlx}^2 \Gamma_I^2$, the number density can be evaluated as  
$n_L(0) \slash n_L(T_\mathrm{rlx}) = \left[N_L(0) \slash N_L(T_\mathrm{rlx})\right](a_\mathrm{rlx}^3 \slash a_R^3)= 
\left[N_L(0) \slash N_L(T_\mathrm{rlx})\right](t_\mathrm{rlx}^2 \Gamma_I^2)$.  This leads to an approximate expression for the asymmetry:  
\begin{align}
 \eta \equiv & \frac{n_L}{(2 \pi^2/45) g_*T^3}  \nonumber \\
 = & \dfrac{45}{2\pi^2} \frac{\sqrt{\lambda}\phi_{0}^{3} \Lambda_I}{M_{n}^{2}T_R^2}\ 
 t_{\textrm{rlx}}^2 \Gamma_I^2 \, \times \, \min \left \{1,  T_{\textrm{rlx}}^{3} t_{\textrm{rlx}} \sigma_{R} \right\} \nonumber
 \\ 
& \times \exp\left[-\left( \frac{24 + 3\sqrt{15}}{\sqrt{ 3 g_* \pi^7}} \right) \sigma_R M_P T_R\right].
\label{eq:eta_analytical}
\end{align}
This analytical estimate agrees within one order of magnitude with the numerical results presented below. 
\begin{figure}[ht!]
\includegraphics[width=1\columnwidth]{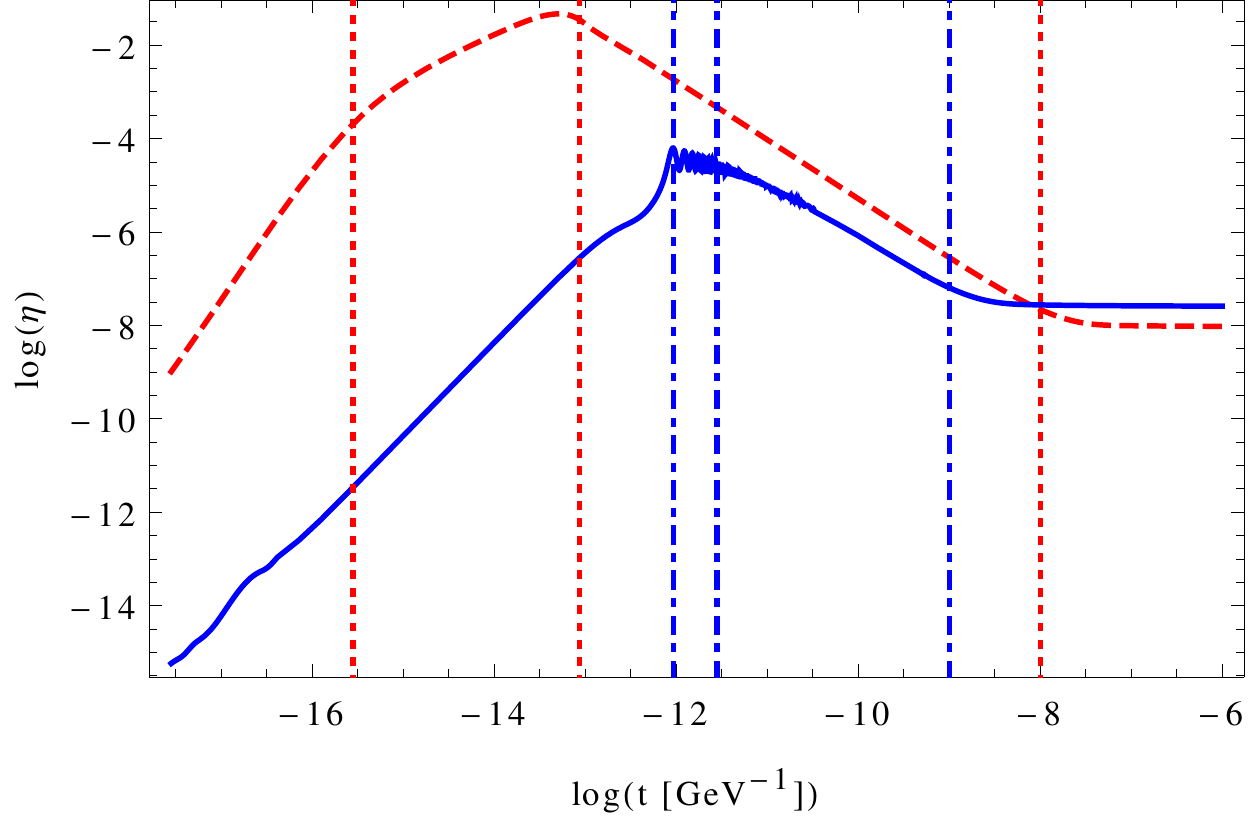}\\
\protect\caption{\label{fig:solution}Numerical solutions for the lepton number asymmetry. 
The IC-1 scenario is shown by the blue solid line for  $\Lambda_{I}=10^{15}$~GeV, $\Gamma_{I}=10^{9}$~GeV, and $M_R = 9 \times 10^{15}$~GeV, which results in $T_{\mathrm{max}} = 6 \times 10^{13}$~GeV, sufficient to destabilize the second minimum.  The initial VEV is $\phi_0 = 1 \times 10^{15}$~GeV.  The IC-2 case is shown by the red dashed line for 
$\Lambda_{I}=10^{17}$~GeV, $\Gamma_{I}=10^{8}$~GeV, $M_{n}=5 \times 10^{12}$~GeV, and $M_{R}=10^{16}$~GeV. The maximum temperature during reheating is $T_{\mathrm{max}}=3\times10^{14}$~GeV, and $\phi_0 = 1 \times 10^{15}$~GeV.  The vertical blue dot-dashed (IC-1) lines denote the first crossing of zero, the time of maximum reheating, and the beginning of the radiation-dominated era, from left to right. The vertical red dotted (IC-2) lines denote the time of maximum reheating, the first crossing of zero, and the beginning of the radiation-dominated era, from left to right.}
\end{figure}

We have analyzed the evolution of the asymmetry numerically. The equation of motion for $\phi(t)$ is 
\begin{equation}
\ddot{\phi} + 3 H(t) \dot{\phi} + V^\prime_\phi \left[\phi,T(t)\right] = 0. 
\label{eq:phioft}
\end{equation}
Here the Hubble parameter is determined by the system of equations 
\begin{align}
H\equiv \frac{\dot a}{a} = \sqrt{ \dfrac{8 \pi}{3 M_P^2} (\rho_r + \rho_I)}, \\
\dot{\rho}_r + 4 H(t) \rho_r = \Gamma_I \rho_I,
\end{align}
where $\rho_I = \Lambda_I^4 e^{-\Gamma_I t} \slash a(t)^3$ is the energy density of the inflaton field. We have included one-loop corrections~\cite{Degrassi:2012ry} and finite temperature corrections~\cite{Kapusta:2006pm} in the Higgs potential.  The solution of Eq.~\eqref{eq:phioft} determines the effective chemical potential $\mu_{\rm eff}(t)$ via Eq.~\eqref{eq:mu}, which we 
then used in solving Eq.~\eqref{eq:nL} numerically.  The results are shown in Fig.~\ref{fig:solution}.

For the IC-1 case shown in Fig.~\ref{fig:solution} by a red dashed line, we set $M_n=T$ in the denominator of Eq.~\eqref{eq:mu}, as expected if the operator $O_6$ is generated by thermal loops.  
For the values of $\Lambda_I$ and $\Gamma_I$ listed in the caption, the effective mass of the Higgs field in the false vacuum is indeed larger than $H_I$, suppressing the baryonic isocurvature perturbations.  For the IC-2 case, shown by the blue solid line, we used a constant value of $M_n$ in Eq.~\eqref{eq:mu}. 
The lepton asymmetries at the end of reheating are $\sim \mathcal{O}(10^{-8})$ in both cases.  
As the Universe cools down, the standard model degrees of freedom go out of equilibrium, and the entropy production reduces the value of the baryon asymmetry by a factor  $\xi \approx 30$.   This brings the final asymmetry to the observed value of $\mathcal{O}(10^{-10})$.  As discussed above, the sphaleron processes redistribute this asymmetry between lepton and baryon numbers, as in the case of thermal leptogenesis~\cite{Fukugita:1986hr}, leading to a successful baryogenesis.

We note that the reheat temperature controls the dilution of $\eta$ in Eq.~\eqref{eq:eta_analytical} via an exponential factor $\exp [-0.036\, \sigma_R M_P T_R]=\exp[-T_R/3\times 10^{13}\:{\rm GeV}]$. 
This implies an upper bound on $T_R \lesssim 3\times 10^{14}$~GeV, to avoid excessive dilution. 

Finally, we note that the epoch of Higgs relaxation can have other observable consequences; in particular, it has the necessary conditions 
for primordial magnetogenesis~\cite{Kandus:2010nw}, although it may be challenging to obtain the large correlation length necessary to explain the seed magnetic fields and the fields 
observed in the voids~\cite{Essey:2010nd}.

In summary, we have shown that the matter-antimatter asymmetry could be generated during Higgs relaxation, assuming that the standard model is valid up to some very high scales. 

The authors thank K.~Harigaya, A.~Kamada, M.~Kawasaki, K.~Schmitz, and T.~T.~Yanagida for the helpful comments and discussions.
This work was supported by the U.S. Department of Energy Grant No.\ DE-SC0009937
and by the World Premier International Research Center Initiative
(WPI Initiative), MEXT, Japan. It was also supported by the National Science Foundation Grant No.\ PHYS-1066293 and the hospitality of the Aspen Center for Physics.

\end{document}